\documentclass[conference]{IEEEtran}
\usepackage{times} 
\usepackage{helvet}  
\usepackage{courier}  
\usepackage[hyphens]{url}  
\usepackage{graphicx}
\usepackage{algorithm}
\usepackage{times}
\usepackage{helvet}
\usepackage{courier}
\usepackage{amssymb}
\usepackage{amsmath}
\usepackage{amsmath, algorithm, algpseudocode}
\usepackage{enumitem}
\usepackage{listings}
\usepackage{xcolor}
\usepackage{booktabs}
\usepackage{multirow}
\usepackage{caption} 
 \usepackage{listings}
\usepackage{xcolor}
\usepackage{tcolorbox}
\usepackage{tcolorbox}
\usepackage[numbers,sort&compress]{natbib} 
\usepackage{caption} 
\tcbuselibrary{listings, skins}

\tcbuselibrary{listings, breakable}

\lstdefinelanguage{json}{
  morestring=[b]",%
  morecomment=[l]{//},
  morekeywords={true,false,null},
  stringstyle=\color{red},
  keywordstyle=\color{blue},
  commentstyle=\color{gray},
  basicstyle=\ttfamily,
  backgroundcolor=\color{gray!15},
  frame = single,
  rulecolor=\color{black}
}
\usepackage{newfloat}
\usepackage{listings}
\DeclareCaptionStyle{ruled}{labelfont=normalfont,labelsep=colon,strut=off} 
\floatstyle{ruled}
\newfloat{listing}{tb}{lst}{}
\floatname{listing}{Listing}
\begin{document}
\title{FlexiDataGen: An Adaptive LLM Framework for Dynamic Semantic Dataset Generation in Sensitive Domains}
\author{
Hamed Jelodar, Samita Bai, Roozbeh Razavi-Far, Ali A. Ghorbani\\
\textit{Canadian Institute for Cybersecurity} \\
\textit{Faculty of Computer Science} \\
\textit{University of New Brunswick} \\
Fredericton, Canada \\
\{h.jelodar, Samita.bai, roozbeh.razavi-far, ghorbani\}@unb.ca
}

\maketitle

\begin{abstract}
Dataset availability and quality remain critical challenges in machine learning, especially in domains where data are scarce, expensive to acquire, or constrained by privacy regulations. Fields such as healthcare, biomedical research, and cybersecurity frequently encounter high data acquisition costs, limited access to annotated data, and the rarity or sensitivity of key events. These issues—collectively referred to as the dataset challenge—hinder the development of accurate and generalizable machine learning models in such high-stakes domains. To address this, we introduce FlexiDataGen, an adaptive large language model (LLM) framework designed for dynamic semantic dataset generation in sensitive domains. FlexiDataGen autonomously synthesizes rich, semantically coherent, and linguistically diverse datasets tailored to specialized fields. The framework integrates four core components: (1) syntactic-semantic analysis, (2) retrieval-augmented generation, (3) dynamic element injection, and (4) iterative paraphrasing with semantic validation. Together, these components ensure the generation of high-quality, domain-relevant data. Experimental results show that FlexiDataGen effectively alleviates data shortages and annotation bottlenecks, enabling scalable and accurate machine learning model development.
\end{abstract}

\section{Introduction}
Natural Language Processing (NLP) has seen rapid progress in recent years, driven largely by advances in large language models (LLMs) and the availability of large, high-quality datasets \cite{kocc2025survey} \cite{jelodar2025large2}. However, in many real-world applications—especially in sensitive or high-stakes domains—access to such datasets remains a major challenge. Fields like healthcare \cite{pais2024overcoming}, biomedical research, and cybersecurity \cite{datta2024emerging} often face barriers such as strict privacy regulations, high data collection costs, and the rarity of critical events \cite{jelodar2025large3}. For example, clinical records require patient consent and ethical oversight, while cybersecurity logs may contain confidential information that cannot be shared.\\

\begin{figure}[h]
    \centering
    \includegraphics[width=1.1\linewidth]{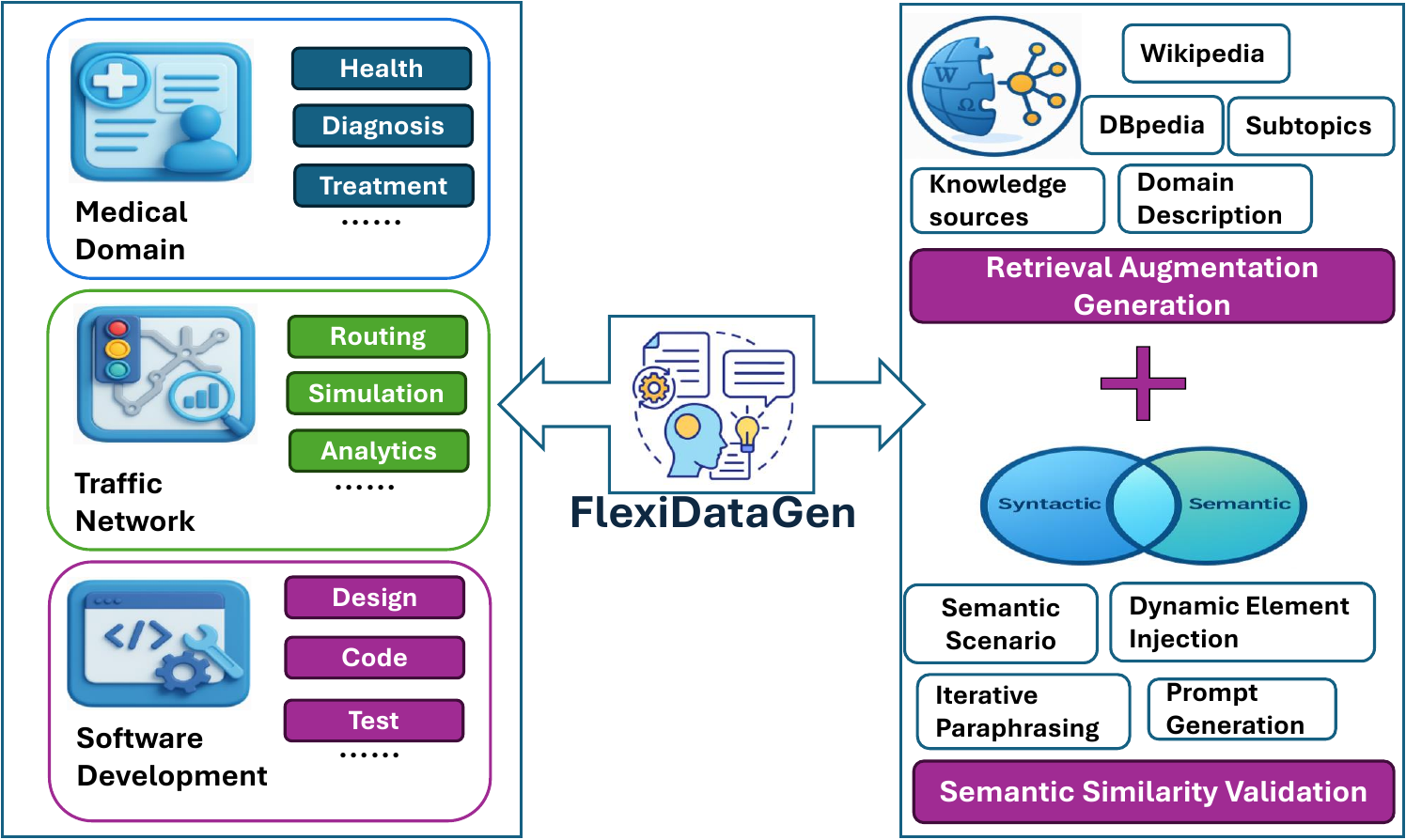}
 \caption{Application scenarios for domain-adaptive data generation using FlexiDataGen}
    \label{fig:Template}
\end{figure}

Large Language Models (LLMs) have transformed natural language processing (NLP) by enabling high-quality language understanding, generation, and reasoning across diverse applications \cite{annepaka2025large, kumar2024large, kukreja2024literature}. However, their effectiveness is highly dependent on access to large-scale, high-quality datasets, especially prompt datasets, for tasks such as prompt tuning, data augmentation, and performance evaluation. Creating such datasets, particularly for sensitive or specialized domains, remains a significant challenge due to linguistic variability, limited domain-specific corpora, and strict privacy constraints.

In this work, we introduce \textbf{FlexiDataGen}, an \textit{adaptive LLM-based framework} for the generation of dynamic semantic datasets in sensitive domains. FlexiDataGen systematically synthesizes semantically coherent, contextually rich, and linguistically diverse prompts tailored to specific domains such as healthcare, biomedical research, and cybersecurity. The framework integrates core techniques, syntactic-semantic analysis, retrieval-augmented generation (RAG) \cite{gao2023retrieval}, dynamic element injection, and iterative paraphrasing with semantic validation to ensure the generation of high-quality prompts, while preserving domain fidelity.  Figure 1 shows an example of an application of the model for different domains.

In this study, we aim to address key limitations in prompt dataset construction by combining domain knowledge retrieval, adaptive scenario definition, and semantic-aware paraphrasing. Through comprehensive experimentation, we demonstrate that this approach enhances the model robustness, generalizability, and applicability in specialized or privacy-sensitive domains. The main contributions of this paper are as follows:

\begin{itemize}
    \item We propose FlexiDataGen, the first modular, adaptive LLM-based framework that automatically generates large-scale, semantically rich prompt datasets across sensitive and domain-specific settings. 

    \item We incorporate retrieval-augmented generation (RAG) into the framework to extract subtopics and contextual descriptions from structured and unstructured knowledge sources such as Wikipedia and DBpedia \cite{hofer2025dbpedia}, thereby enhancing semantic relevance and domain adaptability.

    \item We introduce semantic scenario definition and dynamic prompt instantiation, moving beyond static templates to generate prompts enriched with real-world context, task intent, and domain-specific perspectives.

    \item To the best of our knowledge, this is the first adaptive framework that combines RAG and advanced NLP methods to dynamically generate semantically rich, context-aware prompts for dataset construction, especially in sensitive fields such as healthcare, legal, finance, etc.
\end{itemize}


\section{FlexiDataGen Framework}

The FlexiDataGen framework is a scalable and modular pipeline designed to generate large, diverse, and semantically coherent prompt datasets for sensitive domains, as shown in Figure 2. By decomposing prompt generation into five distinct phases, FlexiDataGen harnesses the power of LLM, RAG, and semantic similarity validation to maximize both dataset quality and contextual relevance.

\begin{figure*}[h]
    \centering
    \includegraphics[width=\linewidth]{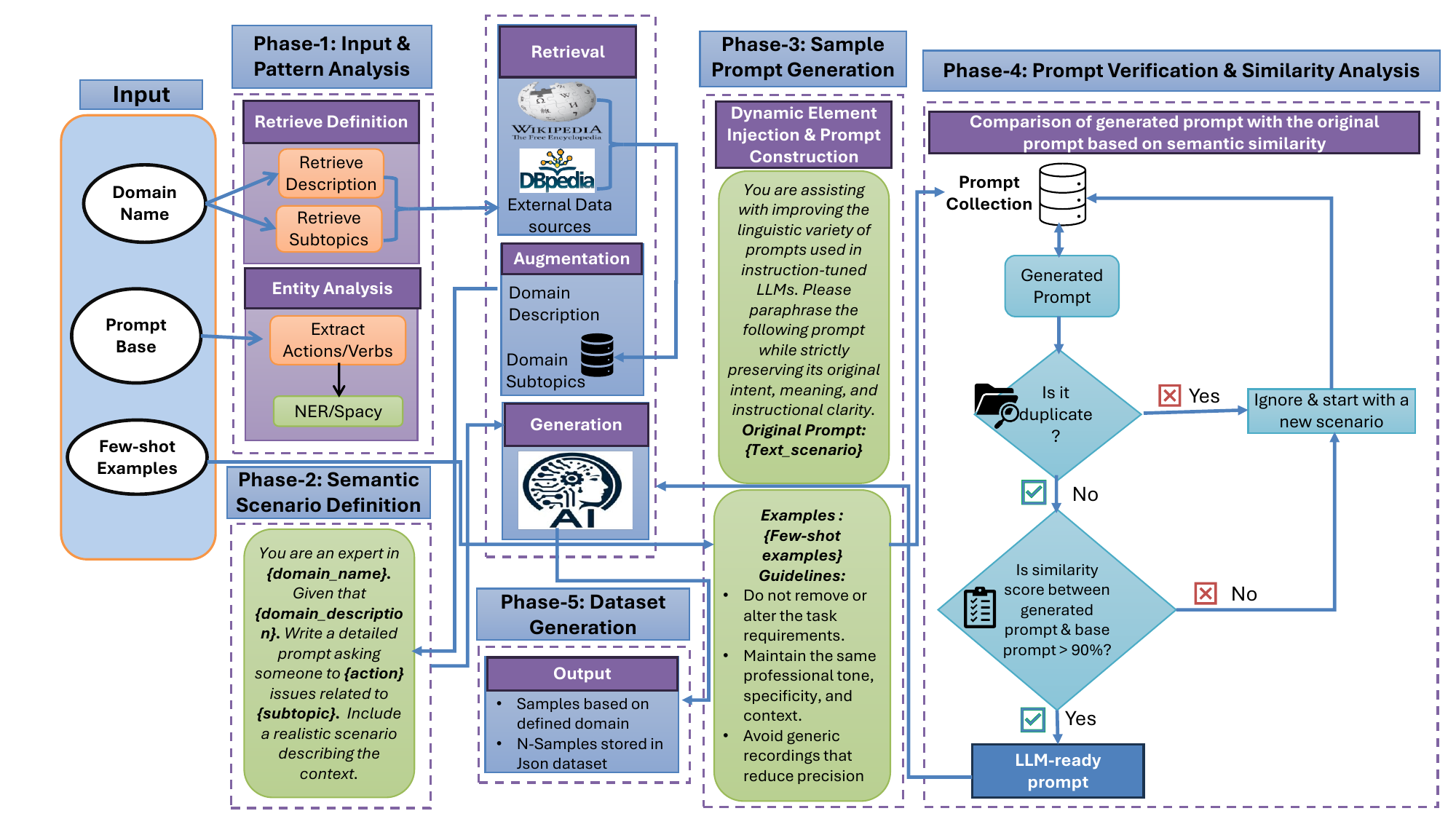}
    \caption{Conceptual Diagram of the Adaptive LLM Framework for Dynamic Semantic Dataset Generation.}
    \label{fig:Template}
\end{figure*}

Each phase builds upon the previous to progressively refine prompts from high-level templates to semantically validated paraphrases suitable for downstream applications such as prompt tuning, data augmentation, or model evaluation.

\subsection{Phase 1: Input and Pattern Analysis}

The first phase begins with a base prompt template \(P\) and a target domain \(D\). The goal is to parse and understand the prompt’s syntactic and semantic structure, extracting key components essential for dynamic prompt generation.\\

Concretely, \(P\) is treated as a parametric template with placeholders \( \{x_i\} \), where \( \{x_i \in X\} \) represent variables, to be filled with meaningful content. Extracting these placeholders enables identification of variable roles. Simultaneously, extracting action verbs \(A = \{a_j\}\) via dependency parsing allows the framework to capture the prompt’s functional intent.\\

Beyond parsing the template, Phase 1 retrieves a domain description \(desc(D)\) using RAG methods from authoritative sources like Wikipedia. This contextual grounding aids the retrieval of relevant subtopics \(S = \{s_k\}\) from structured knowledge bases like DBpedia. This dual syntactic-semantic analysis produces a blueprint tuple:

\begin{equation}
(P, D, X, A, desc(D), S)
\end{equation}

for use in subsequent stages.

\subsection{Phase 2: Semantic Scenario Definition}

Phase 2 defines how prompt elements interact in realistic contexts by crafting semantic scenarios \(C = \{c_l\}\) that represent plausible use cases within domain \(D\). For each subtopic \(s_k \in S\) and action \(a_j \in A\), the framework generates a corresponding set of contextualized scenarios:

\begin{equation}
C_{k,j} = \text{generate\_scenarios}(s_k, a_j)
\end{equation}

These scenarios add situational variability (e.g., urgency, intent, tone) and significantly enrich dataset diversity. Scenarios are generated using pre-trained LLMs and RAG to ensure both creativity and semantic fidelity. Modeling different perspectives, for example, user-focused, outcome-driven, or issue-centered, allows the dataset to reflect a broader range of practical contexts for LLM tuning or evaluation.

\subsection{Phase 3: Dynamic Element Injection and Prompt Construction}

Phase 3 transforms the abstract prompt template into concrete prompt instances. Using selected subtopics \(s_k\) and scenarios \(c_l\), placeholders are dynamically injected as:

\begin{equation}
\text{inject}(P, s_k, c_l) = P \big|_{x_1 := s_k, \, x_2 := c_l}
\end{equation}

This substitution yields unique and meaningful prompts. If a template lacks sufficient placeholders, additional contextual elements are explicitly appended.

To ensure diversity and avoid repetition, the duplication filter \(F\) checks prompt uniqueness:

\begin{equation}
F(p) = 
\begin{cases}
\text{True} & \text{if } p \notin \mathcal{H} \\
\text{False} & \text{otherwise}
\end{cases}
\end{equation}

where \(\mathcal{H}\) tracks the prompt history.

Optionally, prompts can be annotated with metadata (such as domain, tone, and modality) for downstream tasks. These prompt construction and filtering steps form the foundation for the dataset generation process described in Algorithm 1.

\subsection{Phase 4: Iterative Paraphrasing and Semantic Similarity Validation}

To increase linguistic variety and reduce overfitting, Phase 4 applies iterative paraphrasing to prompts \(p\), generating variants \(p'\) using a pre-trained LLM \(M\):

\begin{equation}
p' = M.\text{paraphrase}(p)
\end{equation}

To prevent semantic drift, we compute semantic similarity using an embedding model \(E\):

\begin{equation}
sim(p, p') = \cos \left( E(p), E(p') \right)
\end{equation}

Only paraphrases with similarity above threshold \(\theta\) are accepted:

\begin{equation}
\text{accept}(p, p') = 
\begin{cases}
\text{True} & \text{if } sim(p, p') \geq \theta \\
\text{False} & \text{otherwise}
\end{cases}
\end{equation}

If no valid paraphrase is found after a set number of attempts, the original prompt is retained. Optionally, paraphrases can be clustered to evaluate internal diversity using unsupervised methods. The paraphrasing and validation procedure is detailed in Algorithm 2.

\subsection{Phase 5: Dataset Generation and Output}

In this phase, all previously generated components converge to produce a comprehensive and high-quality prompt dataset ready for use. This final step ensures that the dataset meets desired size and quality criteria through iterative sampling and validation. The final phase composes the complete dataset \(\mathcal{D}\) of target size \(N\):

\begin{equation}
\mathcal{D} = \{ (p_i, p_i') \mid i=1, \ldots, N \}
\end{equation}

Each pair \((p_i, p_i')\) consists of an injected prompt and its semantically validated paraphrase.

The framework iteratively samples domain concepts, injects them into templates (Phase 3), and performs paraphrasing with validation (Phase 4) until the dataset reaches the desired size. This iterative generation loop is summarized in Algorithm 1. \\

However, FlexiDataGen supports continuous learning workflows, allowing new domains \(D_t\) to be introduced using few-shot bootstrapping. Its modularity allows customization of each generation phase to fit new domain or task requirements. Figure 3 shows how the generated prompts interact with the model to create the final dataset.

\begin{figure}[h]
\centering
    \includegraphics[width=\linewidth]{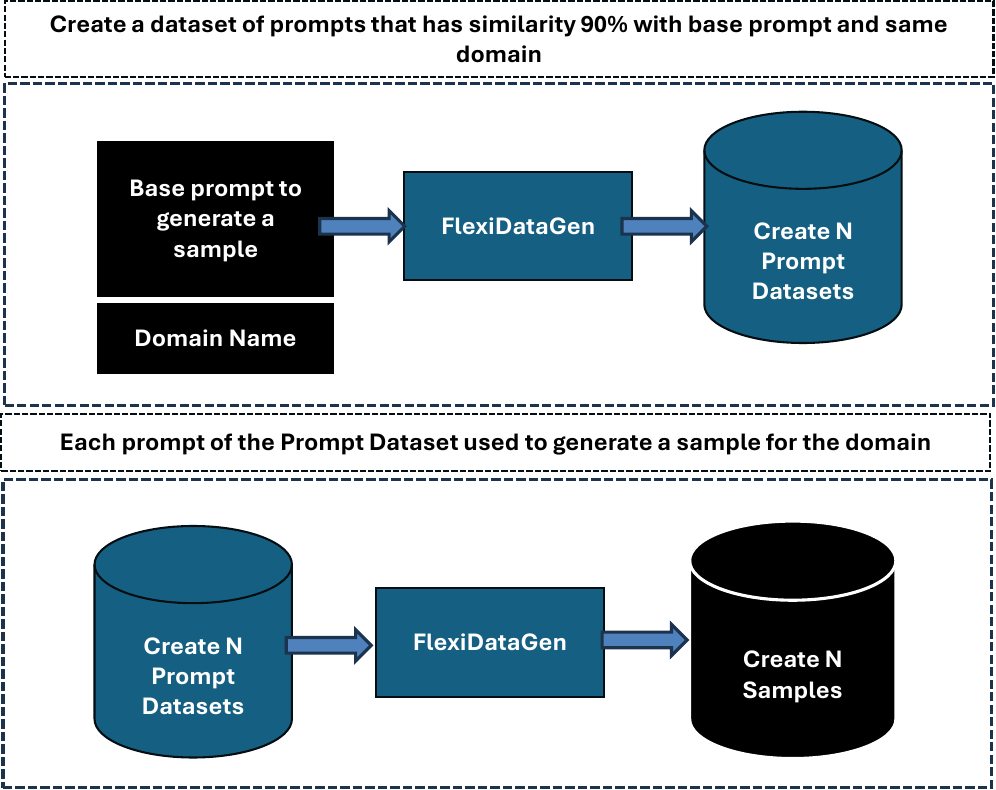}
    \caption{The framework comprises two key components: the first focuses on constructing the prompt dataset, while the second is dedicated to generating the final sample dataset.}
    \label{fig:Template}
\end{figure}

\begin{algorithm}[H]
\caption{Generate FlexiDataGen Prompt Dataset}
\begin{algorithmic}[1]
\Require Base prompt template \(P\), domain \(D\), target dataset size \(N\)
\State Extract placeholders \(X\), actions \(A\) from \(P\)
\State Retrieve domain description \(desc(D)\) via RAG
\State Retrieve subtopics \(S\) related to \(D\) via RAG
\State Initialize dataset \(\mathcal{D} \gets \emptyset\)
\State Initialize prompt history \(\mathcal{H} \gets \emptyset\)
\While{ \(|\mathcal{D}| < N\) }
    \State Sample subtopic \(s_k \sim S\)
    \State Sample action \(a_j \sim A\)
    \State Generate scenarios \(C_{k,j} \gets \text{generate\_scenarios}(s_k, a_j)\)
    \State Sample scenario \(c_l \sim C_{k,j}\)
    \State Construct prompt \(p \gets \text{inject}(P, s_k, c_l)\)
    \If{ \(F(p) = \text{True}\) }
        \State Paraphrase \(p' \gets M.\text{paraphrase}(p)\)
        \If{ \(\text{accept}(p, p') = \text{True}\) }
            \State \(\mathcal{D} \gets \mathcal{D} \cup \{(p, p')\}\)
            \State \(\mathcal{H} \gets \mathcal{H} \cup \{p\}\)
        \EndIf
    \EndIf
\EndWhile
\State \Return \(\mathcal{D}\)
\end{algorithmic}
\end{algorithm}

\begin{algorithm}[H]
\caption{Paraphrase and Validate Prompt}
\begin{algorithmic}[1]
\Require Original prompt \(p\), similarity threshold \(\theta\), max attempts \(T\)
\For{attempt \(t = 1\) to \(T\)}
    \State Generate paraphrase \(p' \gets M.\text{paraphrase}(p)\)
    \State Compute similarity \(s \gets sim(p, p')\)
    \If{ \(s \geq \theta\) }
        \State \Return \(p'\)
    \EndIf
\EndFor
\State \Return \texttt{null} \Comment{Failed to generate valid paraphrase}
\end{algorithmic}
\end{algorithm}

\section{Experiment}

This section details the experimental setup used to evaluate the FlexiDataGen framework. We describe the computing infrastructure, configuration parameters, and hyper-parameter settings used when leveraging large language models (LLMs) for paraphrasing and semantic similarity validation.

\begin{figure*}[h]
    \centering
    \includegraphics[width=0.68\linewidth]{ 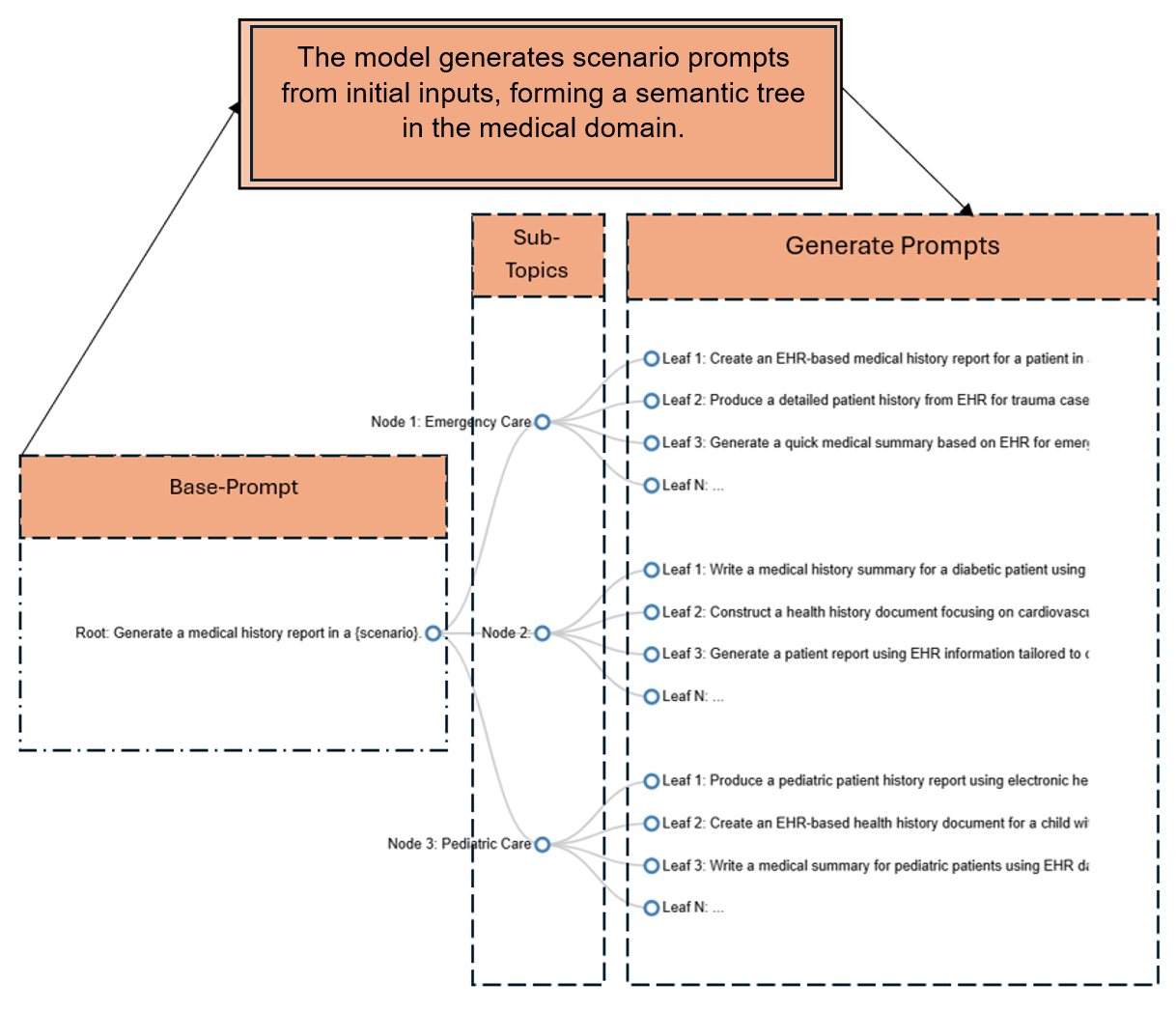}
    \caption{ An Example of Semantic Prompt Tree for EHR (Electronic Health Records) Scenario Generation. }
    \label{fig:Template}
\end{figure*}

\begin{table*}[h]
\centering
\caption{Statistics of Generated Samples: Total, Unique, and Noisy Entries}
\label{tab:raw}
\begin{tabular}{|l|c|c|c|}
\hline
\textbf{Model} & \textbf{\# Generated} & \textbf{Unique} & \textbf{Noisy samples} \\ \hline
Qwen/Qwen2.5-1.5B-Instruct-GPTQ-Int4 & 2000 & 1393 & 606 \\ \hline
microsoft/Phi-4-mini-instruct & 2000 & 1956 & 43 \\ \hline
deepseek-ai/DeepSeek-R1-Distill-Qwen-1.5B & 2000 & 931 & 1038 \\ \hline
meta-llama/Llama-3.2-1B-Instruct & 2000 & 1869 & 130 \\ \hline
\end{tabular}
\end{table*}

\begin{table*}[h]
\centering
\caption{Unique-to-total ratio and noise-to-total ratio.}
\label{tab:ratios}
\begin{tabular}{|l|c|c|}
\hline
\textbf{Model} & \textbf{Unique/Total} & \textbf{Noise/Total} \\ \hline
Qwen/Qwen2.5-1.5B-Instruct-GPTQ-Int4 & 0.697 & 0.303 \\ \hline
microsoft/Phi-4-mini-instruct & 0.978 & 0.022 \\ \hline
deepseek-ai/DeepSeek-R1-Distill-Qwen-1.5B & 0.466 & 0.519 \\ \hline
meta-llama/Llama-3.2-1B-Instruct & 0.934 & 0.065 \\ \hline
\end{tabular}
\end{table*}

\subsection{Computing Infrastructure}

All experiments were conducted in a high-performance computing environment equipped with NVIDIA H100 GPUs. The framework was implemented in Python 3.10 using \texttt{PyTorch 2.0}, the \texttt{Transformers} library \cite{wolf-etal-2020-transformers}, and \texttt{Sentence-Transformers} \cite{reimers-2019-sentence-bert}. Additional components included:

\begin{itemize}
    \item \textbf{LangChain} — to define structured pipelines and modular flows within the framework \cite{langchain}.
    \item \textbf{LlamaIndex} — for Retrieval-Augmented Generation (RAG) using structured and unstructured sources like Wikipedia and DBpedia \cite{llamaindex}.
\end{itemize}

GPU acceleration significantly reduced paraphrasing and inference times, enabling scalable dataset generation. Network access was enabled for real-time querying of external APIs during the RAG process.

\subsection{LLM Hyper-parameters}
The paraphrasing model's parameters were tuned to balance conciseness, diversity, and semantic fidelity. The maximum output length was limited to 512 tokens. A temperature of 0.8 and top-p of 0.9 encouraged diverse yet coherent generation. One paraphrase was generated per attempt, with up to five attempts allowed. Semantic similarity filtering with a threshold of 0.75 ensured faithfulness to the original. All hyperparameters were empirically selected for an optimal trade-off between variability and meaning.







\subsection{Use Cases and Applications}

FlexiDataGen uses domain knowledge, subtopics, and real-life scenarios to create large and varied datasets tailored to specific fields. Here’s an example from the medical domain. The model starts with a basic domain description, such as:

\begin{quote}
\textit{"Healthcare involves the maintenance or improvement of health via diagnosis and treatment."}
\end{quote}

From this starting point, the model automatically identifies related subtopics like \textit{cardiology}, \textit{neurology}, and \textit{pediatrics} by querying external resources such as DBpedia. Next, it generates realistic scenarios for each subtopic—for example, “emergency case with acute symptoms” or “routine wellness checkup.” These elements are then combined into prompt templates, such as:

\begin{quote}
\textit{"Generate a medical history report focusing on \{subtopic\} for a patient in a \{scenario\}."}
\end{quote}

By systematically combining different subtopics and scenarios, FlexiDataGen can produce millions of diverse and contextually rich medical text samples. This capability supports the training of more robust medical NLP models by providing highly relevant and varied data. Figure~4 illustrates how the original prompt expands into many new prompts in a semantic tree structure.

\section{Model Performance in the Medical Use Case }

To assess the practical performance of the FlexiDataGen framework, we tested it with four lightweight instruction-tuned LLMs. Each model generated 2000 prompt samples (Medical domain) for the same template and target domain. The process measured prompt uniqueness and semantic noise filtering. The experimental results summarized in Tables~\ref{tab:raw} and \ref{tab:ratios} reveal several important insights into the performance and robustness of the FlexiDataGen framework. Despite all four models generating the same number of initial samples (2000 prompts), the number of unique prompts and the proportion filtered as noise varied substantially. Notably, the \texttt{microsoft/Phi-4-mini-instruct} \cite{phi4miniinstruct} and \texttt{meta-llama/Llama-3.2-1B-Instruct} models \cite{llama3.2instruct} were capable of generating the highest number of unique samples (1956 and 1869 respectively as shown in Table ~\ref{tab:raw}. These models achieved the highest dataset diversity and lowest noise rates, with the retention rates of 97.8\% and 93.4\% of generated prompts, respectively. This suggests that these models can produce a broad range of semantically coherent paraphrases and scenarios with minimal drift, directly benefiting from the semantic similarity validation phase that preserves quality.

The table \ref{tab:ratios} also shows that the FlexiDataGen modular design successfully filters out incoherent prompts, regardless of the base model. Even when models produced significant noise, the framework ensured that the final dataset maintained only semantically validated and diverse samples. This demonstrates FlexiDataGen's effectiveness in maintaining dataset quality and contextual relevance by balancing creativity with strict semantic filtering.\\

\section{Discussion}

The experimental results demonstrate FlexiDataGen’s effectiveness in generating semantically rich and domain-adapted prompt datasets. Its modular design—relying on domain descriptions, subtopics, and semantic scenarios—enables broad generalization across various fields, including finance, legal technology, customer service, and cybersecurity. The framework supports both general-purpose and specialized use cases, facilitating scalable and privacy-respecting data generation for low-resource settings. Scenario injection further ensures relevance and diversity in the synthetic prompts, boosting robustness during model training and evaluation.

Despite these strengths, FlexiDataGen faces several limitations. First, its performance depends on the quality of input knowledge and scenario construction, which can be challenging in highly specialized or dynamic domains. Second, while paraphrasing supports lexical diversity, it may introduce semantic drift, impacting the realism of generated data. These issues highlight the need for improved knowledge integration and tighter feedback loops with domain experts. \\\\

Future work will focus on enhancing FlexiDataGen by integrating dynamic scenario generation from real-time data, multilingual support, and expert-in-the-loop validation pipelines. Additionally, we plan to explore the controlled use of jailbreak-style prompting techniques \cite{wei2023jailbroken} \cite{jeong2025playing} to elicit hard-to-generate or adversarial cases, especially in security-sensitive domains \cite{cheng2022understand}. These extensions aim to further increase the framework’s adaptability, reliability, and value across emerging NLP applications.

\section{Conclusion}
\label{sec:conclusion}

In this paper, we introduced FlexiDataGen, a dynamic and adaptive LLM framework for semantic dataset generation in sensitive and specialized domains. By combining retrieval-augmented generation, syntactic-semantic decomposition, dynamic prompt injection, and semantically validated paraphrasing, FlexiDataGen enables scalable, high-quality data synthesis across complex fields such as medicine, cybersecurity, and software development.The framework's modularity allows rapid adaptation to new domains with minimal manual intervention, while its semantic validation loop ensures fidelity and diversity in generated content. Experimental results show strong performance in output quality, generalization, and practical utility. FlexiDataGen paves the way for robust, privacy-preserving, and context-aware dataset generation, supporting downstream tasks such as model fine-tuning, evaluation, and domain-specific NLP pipeline development.


\bibliographystyle{ieeetr}
\bibliography{main}

\end{document}